\pdfoutput=1
\documentclass[lettersize,journal]{IEEEtran}
\usepackage{amsmath,amsfonts,amssymb}
\usepackage{amsthm}
\usepackage{algorithmic}
\usepackage{algorithm}
\usepackage{array}
\usepackage{textcomp}
\usepackage{stfloats}
\usepackage{url}
\usepackage{verbatim}
\usepackage{graphicx}
\usepackage{cite}
\usepackage{hyperref}
\usepackage{float}
\usepackage{orcidlink}  % Add this to your preamble
\usepackage{caption}
\usepackage{subcaption}
\usepackage{wrapfig}    % For author biographies
\usepackage{graphicx}

\newtheorem{theorem}{Theorem}
\newtheorem{lemma}{Lemma}

\newtheorem{definition}{Definition}
\newtheorem{proposition}{Proposition}
\newtheorem{remark}{Remark}
\newtheorem{conditions}{Conditions}

\begin{document}

\title{A Novel Parameter-Tying Theorem in Multi-Model Adaptive Systems: Systematic Approach for Efficient Model Selection}

\title{A Novel Parameter-Tying Theorem in Multi-Model Adaptive Systems: Systematic Approach for Efficient Model Selection}

\author{Farid Mafi\textsuperscript{1}\orcidlink{0000-0002-3726-0282}, 
        Ladan Khoshnevisan\textsuperscript{1}\orcidlink{0000-0002-3489-9857}, \textit{Member, IEEE}, 
        Mohammad Pirani\textsuperscript{2}\orcidlink{0000-0003-2677-2140}, \textit{Senior Member, IEEE}, and~
        Amir Khajepour\textsuperscript{1}\orcidlink{0000-0002-1998-6100}, \textit{Senior Member, IEEE}%<-this % stops a space
\thanks{This document is the results of the research project funded by the Natural Sciences and Engineering Research Council of Canada (NSERC) under Grant ALLRP 566320 – 21.}% <-this % stops a space
\thanks{\textsuperscript{1}F. Mafi, L. Khoshnevisan, and A. Khajepour are with the Department of Mechanical and Mechatronics Engineering, University of Waterloo, Waterloo, ON, N2L 3G1, Canada (e-mail: fmafisho@uwaterloo.ca; lkhoshnevisan@uwaterloo.ca; a.khajepour@uwaterloo.ca).}%
\thanks{\textsuperscript{2}M. Pirani is with the Department of Mechanical Engineering, University of Ottawa, Ottawa, ON, K1N 6N5, Canada (e-mail: mpirani@uottawa.ca).}%
}

% The paper headers
%\markboth{IEEE TRANSACTIONS ON on Automatic Control}%
%{Mafi \MakeLowercase{\textit{et al.}}: A Novel Parameter-Tying Theorem in Multi-Model Adaptive Systems}

%\IEEEpubid{0000--0000/00\$00.00~\copyright~2024 IEEE}

\maketitle

\begin{abstract}
This paper presents a novel theoretical framework for reducing the computational complexity of multi-model adaptive control/estimation systems through systematic transformation to controllable canonical form. While traditional multi-model approaches face exponential growth in computational demands with increasing system dimension, we introduce a parameter-tying theorem that enables significant dimension reduction through careful analysis of system characteristics in canonical form. The approach leverages monotonicity properties and coordinated parameter relationships to establish minimal sets of identification models while preserving system stability and performance. We develop rigorous criteria for verifying plant inclusion within the convex hull of identification models and derive weight transformation relationships that maintain system properties across coordinate transformations. The effectiveness of the framework is demonstrated through application to coupled lateral-roll vehicle dynamics, where the dimension reduction enables real-time implementation while maintaining estimation accuracy. The results show that the proposed transformation approach can achieve comparable performance to conventional methods while requiring substantially fewer identification models, enabling practical deployment in high-dimensional systems.
\end{abstract}


\IEEEpubidadjcol

\section{Introduction}

\subsection{Motivation}
Multi-model adaptive systems has emerged as a powerful framework for handling parametric uncertainties in dynamic systems \cite{RN1,RN3}. These approaches have demonstrated significant advantages over traditional adaptive and robust control methods, particularly in addressing broader uncertainties and overcoming system-theoretic limitations in parameter identification \cite{RN2}. However, as system complexity increases, the "curse of dimensionality" presents a fundamental challenge - the number of required identification models grows exponentially with the system order \cite{RN4}. This computational burden has severely limited the practical application of multi-model methods in high-dimensional systems.

The computational challenge becomes particularly acute in modern applications such as integrated vehicle dynamics \cite{hajiloo2022integrated} and power networks \cite{RN45,RN46}, where multiple subsystems interact through coupled dynamics. This motivates our investigation of systematic dimension reduction through coordinate transformations.

A key insight driving this work is that many physical systems exhibit special structural properties and inherent parameter dependencies that can be exploited to reduce the effective dimension of the uncertainty space. This observation suggests that by carefully analyzing system properties in canonical coordinates, we may be able to achieve significant computational savings while preserving the essential features needed for multi-model adaptive systems.

\subsection{Literature Review}
Handling uncertainties in system modeling remains a critical challenge in dynamic systems. Traditional approaches in control systems such as adaptive control and robust control have been extensively studied, yet they possess inherent limitations in handling systems with broader uncertainties \cite{RN1}, including several system-theoretic limitations for parameter identification and potential system instability \cite{RN3,RN2}. For dynamic systems, particularly those exhibiting time-varying behavior, the multi-model approach has emerged as a promising method for addressing uncertainties \cite{RN4}. Unlike traditional methods constrained by a single model, this approach harnesses multiple models to provide a comprehensive understanding of system behavior across diverse operational conditions \cite{RN5}.

The multi-model approach can be structured in two primary ways: switching and blending. Switching-based schemes focus on selecting one model at a time based on how well it matches the plant's current state, while blending-based approaches combine contributions from all models with specific weights to achieve an overall solution. In switching-based schemes, innovative approaches in adaptive control systems have been extensively studied \cite{RN14}, particularly focusing on the integration of switching mechanisms between models. These schemes introduce two distinct strategies: direct switching, involving real-time transitioning to each controller according to system output \cite{RN15,RN16}, and indirect switching, which determines both the timing and selection of subsequent controllers dynamically \cite{RN17}. Building on these foundations, researchers have proposed supervisory controllers to oversee seamless transitions between candidate set-point controllers \cite{RN18,RN20,Morse}.

The development of blending-based approaches was motivated by two key challenges faced in switching methods. First, as one predetermined model must closely match the original system's parameters at any given time, the quantity of fixed models can increase exponentially with the dimension of the unknown parameter vector \cite{RN32}. Second, apart from the best-fitting model, other models do not contribute to parameter estimation, leading to underutilization of available data. To address these limitations, researchers introduced "second-level adaptation" for identifying plants using fixed identification models \cite{RN5}. This technique was formulated using the indirect model reference adaptive control scheme and later extended to regulate uncertain Linear Time-Variant (LTV) SISO systems in companion form. Subsequent research has applied similar blending concepts to analyze vehicle motion stability \cite{RN31,RN11,RN10}.

The transformation of systems to canonical forms offers potential advantages in revealing inherent parameter dependencies and structural properties. However, the intersection of these transformations with multi-model control theory, especially in the context of dimension reduction and computational efficiency, remains unexplored. This gap in the literature, combined with the pressing need for efficient implementations in high-dimensional systems, motivates our present work on developing a systematic framework for dimension reduction through canonical transformations in multi-model control.

\subsection{Contributions}
This paper makes several novel contributions to the theory and practice of multi-model adaptive systems:

\begin{itemize}
\item Development of a parameter-tying theorem that establishes conditions under which the uncertainty space dimension can be reduced through transformation to controllable canonical form.

\item Derivation of explicit weight transformation relationships that preserve convex hull properties and stability guarantees across coordinate transformations.

\item Introduction of verifiable criteria for ensuring plant inclusion within the convex hull of identification models in canonical coordinates.

\item Characterization of coordinated parameter relationships that enable further reduction in required identification models while maintaining coverage of the uncertainty space.

\item Validation of the theoretical framework through application to coupled lateral-roll vehicle dynamics, demonstrating significant computational savings while maintaining estimation performance.
\end{itemize}

\section{Main Results}
This section presents the paper's main contribution: a new theorem in Multi-Model Adaptive Systems (MMAS) framework. Following the problem statement and preliminaries, we discuss our key findings. The subsequent sections provide a detailed proof of the theorem and analyze its conditions.

\subsection{Problem statement and preliminaries}
Consider the following uncertain linear system in state space form:
\begin{equation}
\dot{x}_p(t) = A_p(\mathbf{m}) x_p(t) + B_p(\mathbf{m}) d(t), 
\label{eq1}
\end{equation}
where:
\begin{itemize}
    \item \( x_p(t) \in \mathbb{R}^n \) is the state vector,
    \item \( d(t) \in \mathbb{R}^h \) is the input vector,
    \item \( A_p(\mathbf{m}) \in \mathbb{R}^{n \times n} \) is the state matrix, 
    \item \( B_p(\mathbf{m}) \in \mathbb{R}^{n \times h} \) is the input matrix,
    \item \( \mathbf{m} \in \mathbb{R}^{k} \) is the uncertain system parameter vector.
\end{itemize}

According to the multi-model adaptive control theory \cite{RN5}, if the plant parameter \( A_p \) lies in the convex hull of \( N \) known identification models of \( \{A_1, A_2, \dots, A_N\} \) for \( t \geq t_0 \), given the fact that \( e_i(t_0) =0 \) for $i \in \{1,\ldots,N\}$, where \(e_i(t) = x_i(t) - x_p(t)\) is defined as the identification error, then it follows that:
\begin{equation}
A_p(t) = \sum_{i=1}^N w_i(t) A_i, \\
\sum_{i=1}^N w_i(t) e_i(t) = 0, 
\label{eq2}
\end{equation}
where \( w_i > 0 \) and \( \sum_{i=1}^N w_i = 1 \).

\begin{definition}\label{def:vector}
A parameter uncertainty vector is defined as $\mathbf{m} = (m_1, m_2, \ldots, m_k) \in \mathbb{R}^k$ where each component $m_i$ is bounded such that:
\begin{equation}
m_i \in [p_i, q_i] \quad \text{for all } i \in \{1,\ldots,k\},
\label{eq3}
\end{equation}
with $p_i, q_i \in \mathbb{R}$ and $p_i < q_i$. The set of all possible parameter uncertainty vectors forms a bounded subset $M \subset \mathbb{R}^k$ given by:
\begin{equation}
M = [p_1, q_1] \times [p_2, q_2] \times \cdots \times [p_k, q_k],
\label{eq4}
\end{equation}
where '$\times$' denotes the Cartesian product. For the uncertain linear system, the state matrix $A_p(\mathbf{m}) = [a_{ij}(\mathbf{m})]_{n \times n}$ is an $n \times n$ matrix where each entry $a_{ij}: M \to \mathbb{R}$ is a continuous function mapping from the parameter space $M$ to the real numbers.
\end{definition}

\begin{definition}
For the uncertain system $\Sigma_p(t)$ characterized by $(A_p, B_p)$, there exists a non-singular transformation matrix $T_p(t) \in \mathbb{R}^{n \times n}$ that transforms the system into its controllable canonical form (also known as controller companion form):
\begin{equation}
\begin{aligned}
\bar{A}_p(t) &= T_p(t) A_p T_p^{-1}(t) \in \mathbb{R}^{n \times n}, \\
\bar{B}_p(t) &= T_p(t) B_p \in \mathbb{R}^{n \times h},
\end{aligned}
\label{eq5}
\end{equation}
where $\bar{A}_p(t)$ is in companion form with its last row containing the coefficients $[-\alpha_{p,1}(t) \quad -\alpha_{p,2}(t) \quad \cdots \quad -\alpha_{p,n}(t)]$, and each column of $\bar{B}_p(t)$ is $[0 \quad 0 \quad \cdots \quad 1]^T \in \mathbb{R}^n$.

Similarly, for each vertex system $\Sigma_i$, $i \in \{1,\ldots,N\}$, described by the pair $(A_i, B_i)$, there exists a non-singular transformation matrix $T_i \in \mathbb{R}^{n \times n}$ such that:
\begin{equation}
\begin{aligned}
\bar{A}_i &= T_i A_i T_i^{-1} \in \mathbb{R}^{n \times n}, \\
\bar{B}_i &= T_i B_i \in \mathbb{R}^{n \times h},
\end{aligned}
\label{eq6}
\end{equation}
where $\bar{A}_i$ is in companion form with its last row containing the coefficients $[-\alpha_{i,1} \quad -\alpha_{i,2} \quad \cdots \quad -\alpha_{i,n}]$, and each column of $\bar{B}_i$ is $[0 \quad 0 \quad \cdots \quad 1]^T \in \mathbb{R}^n$.
\end{definition}

\begin{definition}\label{def:space}  
Let the vector \( \theta \in \mathbb{R}^{n} \) be defined as:
\begin{equation}
\theta = \begin{bmatrix} \alpha_{1} & \alpha_{2} & \cdots & \alpha_{n} \end{bmatrix}^T,
\label{eq7}
\end{equation}
which contains all elements of last row of the matrix \(\bar{A} \). Let Conv(·) denote the convex hull operator, which for a finite set of points S returns the smallest convex set containing S, defined as:
\begin{equation}
\text{Conv}(S) = \left\{ \sum_{i=1}^{k} \lambda_i x_i \;\middle|\; \{x_i\}_{i=1}^k \subseteq S, \lambda_i \geq 0, 
\sum_{i=1}^{k} \lambda_i = 1 \right\}.
\end{equation}
We define the \textit{Parameter Uncertainty Space} \( \Theta \) as the convex hull of a finite set of admissible points \( \{ \theta^{(1)}, \theta^{(2)}, \dots, \theta^{(C)} \}\) such that:
\begin{equation}
\begin{aligned}
\Theta = \left\{ \Theta \subseteq \mathbb{R}^{n} \;\middle|\; 
\begin{array}{l}
\Theta \triangleq \text{Conv} \left( \{ \theta^{(1)}, \theta^{(2)}, \dots, \theta^{(C)} \} \right), \\
\theta_{i,\text{min}} \leq \theta_i^{(c)} \leq \theta_{i,\text{max}}, \\
\forall i \in \{1, \dots, n\}, \ \forall c \in \{1, \dots, C\}.
\end{array}
\right\},
\end{aligned}
\label{eq8}
\end{equation}
where \( \theta_{i,\text{min}} \) and \( \theta_{i,\text{max}} \) are the lower and upper bounds for the \( i \)-th component of \( \theta \), respectively.
\end{definition}

\subsection{The Parameter-Tying Theorem}
We present a key theorem within the multi-model adaptive systems framework that provides criteria for selecting the identification models. This theorem provides a systematic approach to maximize the parameter uncertainty space coverage while minimizing the number of required identification models.

\begin{theorem}\label{theorem:monotonicity}
Let $\Sigma_p$ be the uncertain linear system defined in \eqref{eq1} with parameter uncertainty vector $\mathbf{m} \in M \subset \mathbb{R}^k$. Define $\Pi_l: M \to M \setminus \{m_l\}$ as the projection operator removing the $l$-th component, and let $\hat{\mathbf{m}}_l = \Pi_l(\mathbf{m})$. Suppose that for each $l \in \{1, 2, \ldots, k\}$ and any fixed $\hat{\mathbf{m}}_l$, the continuous function $g_{l,ij}: [p_l, q_l] \to \mathbb{R}$ given by
    \begin{equation}
        g_{l,ij}(m_l) = a_{ij}(\hat{\mathbf{m}}_l, m_l)
        \label{eq9}
    \end{equation}
    is monotonic $\forall i,j \in \{1,\ldots,n\}$, where $a_{ij}$ are the entries of $A_p(\mathbf{m})$.

Then, the set of identification models required to span the entire uncertain parameter space in controllable canonical form can be constructed by selecting only the boundary values $p_l$ and $q_l$ under the weight transformation \(\tilde{w}_i = w_i S_i \), where the matrix \( S_i = T_p T_i^{-1} \).
\end{theorem}

\subsection{Proof of the Parameter-Tying Theorem}
The proof proceeds by establishing three key properties:
\begin{enumerate}
    \item Monotonicity preservation under parameter variation
    \item Characteristic Polynomial Analysis
    \item Weight Transformation Analysis
\end{enumerate}

Each property is first established individually, and then combined to complete the proof.

\subsubsection{Monotonicity Analysis}

Under the given monotonicity conditions in Theorem~\ref{theorem:monotonicity}, the extremal values of the system matrix elements must occur at the parameter space vertices. We proceed through several lemmas to build this result rigorously.

\begin{lemma}[Single Parameter Extremality]
For any fixed $\hat{\mathbf{m}}_l \in M \setminus \{m_l\}$ and any $i,j \in \{1,\ldots,n\}$, the continuous function $g_{l,ij}(m_l) = a_{ij}(\hat{\mathbf{m}}_l, m_l)$ achieves its extremal values at the boundary points $\{p_l, q_l\}$. This follows from the properties of continuous monotonic functions on closed intervals \cite{rudin1964principles}.
\end{lemma}

\begin{lemma}[Extension to Multiple Parameters]
For any subset of indices $L \subseteq \{1,\ldots,k\}$ and fixed values of parameters not in L, the extremal values of $a_{ij}$ occur when all parameters $m_l$ with $l \in L$ take their boundary values.
\end{lemma}

\begin{proof}
We proceed by induction on $|L|$. The base case $|L| = 1$ is established by Lemma 1. For the inductive step, assume the statement holds for all sets of size $n \geq 1$ and consider a set $L$ of size $n+1$. Let $l \in L$ and $L' = L \setminus \{l\}$. By our inductive hypothesis, for the set $L'$ of size $n$, the extremal values occur at boundary points. Therefore:
\begin{equation}
\min_{M_l \times M_s} a_{ij}(\mathbf{m}) = \min_{M_l \times \{p_s,q_s\}} a_{ij}(\mathbf{m}) = \min_{\{p_l,q_l\} \times \{p_s,q_s\}} a_{ij}(\mathbf{m})
\end{equation}
where $M_l = [p_l,q_l]$, $M_s = [p_s,q_s]$ for $s \in L'$. The first equality holds because for any value in $M_l$, the minimum over $M_s$ occurs at the boundary points $\{p_s,q_s\}$ by our inductive assumption on $L'$. The second equality follows from Lemma 1 applied to the minimization over $M_l$, since for any fixed choice from $\{p_s,q_s\}$, we have a monotonic function over $M_l$. An analogous argument holds for maximization.
\end{proof}

\begin{theorem}[Global Extremality]
For all $i,j \in \{1,\ldots,n\}$:
\begin{align*}
\min\{a_{ij}(\mathbf{m}) \mid \mathbf{m} \in M\} &= \min\{a_{ij}(\mathbf{m}) \mid \mathbf{m} \in \text{vert}(M)\}, \\
\max\{a_{ij}(\mathbf{m}) \mid \mathbf{m} \in M\} &= \max\{a_{ij}(\mathbf{m}) \mid \mathbf{m} \in \text{vert}(M)\},
\end{align*}
where $\text{vert}(M)$ denotes the set of vertices of $M$.
\end{theorem}

\begin{proof}
Apply Lemma 2 with $L = \{1,\ldots,k\}$. Since $M = [p_1, q_1] \times \cdots \times [p_k, q_k]$, its vertices are precisely the points where each parameter takes either its minimum or maximum value.
\end{proof}

This sequence of results establishes that, under the monotonicity conditions of the theorem, we can independently sweep the matrix elements $a_{ij}(s)$ across their lower and upper bounds.

\subsubsection{Characteristic Polynomial Analysis}
Let $A \in \mathbb{R}^{n \times n}$ be a square matrix with elements $a_{ij}$, where each element is bounded:
\begin{equation}
a_{ij} \in [lb_{ij}, ub_{ij}], \quad \forall i,j \in \{1,\ldots,n\}.
\label{eq11}
\end{equation}

The characteristic polynomial $p_A(s)$ of matrix $A$ is defined as:
\begin{equation}
p_A(s) = \det(sI_n - A) = s^n + \sum_{k=0}^{n-1} c_k s^k,
\label{eq12}
\end{equation}
where each coefficient $c_k$ is a function of the matrix elements $a_{ij}$. A detailed analysis of these coefficients (see Appendix \ref{app:char-poly}) establishes that:

\begin{enumerate}
    \item The leading coefficient of $s^n$ is always 1, independent of matrix elements \cite{horn2012matrix}.
    \item The coefficient $c_{n-1}$ equals the negative trace of matrix $A$ and is bounded as:
    \begin{equation}
    -\sum_{i=1}^n ub_{ii} \leq c_{n-1} \leq -\sum_{i=1}^n lb_{ii}
    \end{equation}
    (Appendix \ref{app:n-1-coef})
    \item The constant term $c_0 = (-1)^n \det(A)$ has bounds determined by the Leibniz formula (Appendix \ref{app:constant-term})
    \item Intermediate coefficients $c_k$ ($1 \leq k \leq n-2$) are related to sums of principal minors with bounds established through determinant analysis (Appendix \ref{app:intermediate-coef})
\end{enumerate}

This analysis demonstrates that all coefficients of the characteristic polynomial are constrained by the element-wise bounds of matrix $A$, establishing a direct relationship between parameter uncertainty bounds and system dynamics in canonical form.

\subsubsection{Weight Transformation Analysis}\label{subsec:weight-transform}

The multi-model representation assumes that the uncertain plant matrices can be expressed as convex combinations of known vertex system matrices:

\begin{equation}
\label{eq33}
A_p = \sum_{i=1}^{N} w_iA_i,\quad B_p = \sum_{i=1}^N w_iB_i
\end{equation}

where the scalar weights satisfy standard convexity conditions:
\begin{equation}
\label{eq34}
\sum_{i=1}^{N} w_i = 1,\quad w_i \geq 0,\quad \forall i \in \{1,\dots,N\}
\end{equation}

When transforming to controllable canonical form, we demonstrate in Appendix \ref{app:weight-transform} that this convex combination relationship can be preserved through a matrix-valued weight transformation:

\begin{equation}
\tilde{w}_i = w_i S_i
\end{equation}

where $S_i = T_p T_i^{-1}$ represents the similarity transformation between the uncertain system and vertex system $i$ coordinates. Through substitution of the transformation relationships, we obtain the key result:

\begin{equation}
\label{eq52}
\bar{A}_p = \sum_{i=1}^{N} w_i \left[ \left( \sum_{j=1}^{N} w_j T_j^{-1} \right)^{-1} T_i^{-1} \bar{A}_i T_i \left( \sum_{j=1}^{N} w_j T_j^{-1} \right) \right]
\end{equation}

This formulation establishes that the canonical form of the uncertain system is itself a convex combination of similarity-transformed vertex systems, with the transformation matrices dependent on the convex weights. The practical implementation requires only the determination of scalar weights $w_i$ from plant identification errors in the original coordinate system, followed by the application of these transformations for closed-loop analysis.

The complete derivation of transformation relationships and proof of weight properties is provided in Appendix \ref{app:weight-transform}.

\subsubsection{Completion of the Proof}
We now combine the results from the previous steps:
\begin{enumerate}
    \item From Step 1 (Monotonicity Analysis), we established that under the monotonicity conditions specified in the theorem for functions $g_{l,ij}(m_l)$, the extremal values of all matrix elements $a_{ij}(\mathbf{m})$ occur at the parameter space vertices. This means selecting only the boundary values $p_l$ or $q_l$ for each $l \in \{1, 2, \ldots, k\}$ is sufficient.
    
    \item Step 2 (Characteristic Polynomial Analysis) proved that the coefficients of the characteristic polynomial in the companion form are bounded by functions of the matrix elements' extremal values. Consequently, the entire reachable set of characteristic polynomial coefficients can be represented as convex combinations of the coefficients corresponding to vertex systems.
    
    \item Step 3 (Weight Transformation Analysis) demonstrated that through the transformation $\tilde{w}_i = w_i S_i$, where $S_i = T_p T_i^{-1}$, any system matrix within the parameter uncertainty space can be expressed as a convex combination of the vertex system matrices in companion form, while preserving the companion form structure.
\end{enumerate}

These results establish both the minimality and completeness of the identification model set: the set is minimal because removing any vertex system would create a gap in the parameter uncertainty space that cannot be spanned by the remaining vertices, and it is complete because the parameter uncertainty space is fully characterized by the bounds of the characteristic polynomial coefficients, which are determined by the extremal values at the boundaries. This completes the proof.~$\square$

\subsection{Expanding the Conditions of Parameter-Tying Theorem}

The Parameter-Tying Theorem was established under the condition that each function $g_{l,ij}(m_l)$ is monotonic. We now explore conditions under which pairs of parameter-element functions achieve their extrema at the same parameter values, even if these values are not at the parameter boundaries.

\begin{definition}
For the uncertain system matrix $A_p(\mathbf{m})$, consider two parameter-element functions $g_{l,ij}, g_{l,pq}: [p_l, q_l] \to \mathbb{R}$ for fixed indices $i,j,p,q$ and parameter index $l$, defined as:
\begin{equation}
\begin{aligned}
g_{l,ij}(m_l) &= a_{ij}(\Pi_l^{-1}(\hat{\mathbf{m}}_l, m_l)) \\
g_{l,pq}(m_l) &= a_{pq}(\Pi_l^{-1}(\hat{\mathbf{m}}_l, m_l))
\end{aligned}
\label{eq53}
\end{equation}
where $\Pi_l$ and $\hat{\mathbf{m}}_l$ are as defined previously. These functions are said to be coordinated if they achieve their global extrema at the same parameter values.
\end{definition}

\begin{conditions}
Given two continuously differentiable parameter-element functions $g_{l,ij}$ and $g_{l,pq}$ defined on $[p_l, q_l]$, their global extrema occur at the same parameter values $m_l^*$ if they satisfy any of the following independent conditions:

1) \textbf{Affine Relationship}: There exist constants $\alpha \neq 0$ and $\beta$ such that:
   \begin{equation}
   g_{l,pq}(m_l) = \alpha g_{l,ij}(m_l) + \beta
   \label{eq54}
   \end{equation}
   When $\alpha > 0$, the functions achieve their respective maxima and minima at the same points; when $\alpha < 0$, the maximum of one function corresponds to the minimum of the other.

2) \textbf{Functional Relationship}: There exists a strictly monotonic and continuously differentiable function $h: \mathbb{R} \to \mathbb{R}$ such that:
   \begin{equation}
   g_{l,pq}(m_l) = h(g_{l,ij}(m_l))
   \label{eq55}
   \end{equation}
   The monotonicity of $h$ ensures preservation of extremum points, with $h'(x) \neq 0$ for all $x$ in the range of $g_{l,ij}$.

3) \textbf{Symmetry}: Both functions are symmetric about the same point $m_l^0 \in [p_l, q_l]$ such that:
   \begin{equation}
   \begin{aligned}
   g_{l,ij}(m_l^0 + \delta) &= g_{l,ij}(m_l^0 - \delta) \\
   g_{l,pq}(m_l^0 + \delta) &= g_{l,pq}(m_l^0 - \delta)
   \end{aligned}
   \label{eq56}
   \end{equation}
   for all $\delta \in [-\min(m_l^0-p_l, q_l-m_l^0), \min(m_l^0-p_l, q_l-m_l^0)]$.

4) \textbf{Critical Points}: The functions share critical points with consistent nature:
   \begin{equation}
   \begin{aligned}
   g_{l,ij}'(m_l^*) = 0 &\iff g_{l,pq}'(m_l^*) = 0 \\
   \text{sign}(g_{l,ij}''(m_l^*)) \cdot \text{sign}(g_{l,pq}''(m_l^*)) &= 1
   \end{aligned}
   \label{eq57}
   \end{equation}
   where $m_l^*$ are all critical points in $(p_l, q_l)$, and the second condition ensures both functions have the same type of extremum (maximum or minimum) at each critical point.

5) \textbf{Periodic Structure}: Both functions are periodic with period $T > 0$ and phase difference $\phi \in [0, T)$:
   \begin{equation}
   \begin{aligned}
   g_{l,ij}(m_l + T) &= g_{l,ij}(m_l) \\
   g_{l,pq}(m_l) &= g_{l,ij}(m_l + \phi)
   \end{aligned}
   \label{eq58}
   \end{equation}
   for all $m_l \in [p_l, q_l-T]$, ensuring synchronized extrema with fixed phase offset.
\end{conditions}

\begin{remark}
The conditions above are independent in the sense that no condition can be derived from a combination of others. This independence is established through the following properties:
\begin{itemize}
    \item The Affine Relationship is the only condition that preserves exact linear relationships.
    \item The Functional Relationship allows for nonlinear monotonic transformations not captured by other conditions.
    \item The Symmetry condition can hold for non-periodic, non-monotonic functions.
    \item The Critical Points condition allows for more general relationships between extrema.
    \item The Periodic Structure captures cyclic behaviors not implied by other conditions.
\end{itemize}
\end{remark}

\begin{remark}
Unlike the original Parameter-Tying Theorem where monotonicity ensures extrema occur at parameter space vertices, under these extended conditions, the extrema of parameter-element functions may occur at interior points of the parameter interval. This extension allows for more flexible system representations while maintaining the structured uncertainty framework necessary for multi-model adaptive systems.
\end{remark}

\subsection{A Criterion for Verifying Plant Inclusion in the Convex Combination of Identification Models}
The weight transformation to canonical form presented in Section~\ref{subsec:weight-transform} relies on a crucial assumption: the uncertain plant must lie within the convex hull of the identification models. While our systematic model selection approach typically ensures this condition is met, certain dynamic systems may temporarily violate the uncertainty bounds. We therefore require a rigorous method to detect when the plant deviates from the convex hull of identification models. In this section, we introduce a novel criterion based on the analysis of identification error signs across the model set. This criterion provides both a theoretical foundation for convex hull containment verification and a practical measure for real-time monitoring of model set adequacy.

\begin{theorem}\label{theorem:plant_inclusion}
Let \( E(t) = [e_1(t) \ e_2(t) \ \dots \ e_N(t)] \) be the error vector at time \( t \), where \(e_i(t) = x_i(t) - x_p(t)\). The plant matrices $A_p(\mathbf{m})$ and $B_p(\mathbf{m})$ lie within the convex hull formed by the set $[A_i \ B_i]: i = 1, \dots, N$ if and only if there exist at least two elements of \( E(t) \) with opposite signs.
\end{theorem}

The complete proof of Theorem~\ref{theorem:plant_inclusion}, encompassing both forward and backward directions, is provided in Appendix~\ref{app:plant_inclusion}. The proof demonstrates that the existence of opposite signs in the error vector is both necessary and sufficient for establishing plant inclusion within the convex hull of identification models.

This theorem provides a practical criterion for verifying whether the plant matrices $A_p(\mathbf{m})$ and $B_p(\mathbf{m})$ lie within the convex hull of identification models by monitoring the error vector $E(t)$. This is particularly significant as it enables real-time verification of the fundamental assumption required for the Parameter-Tying Theorem and the subsequent weight transformation analysis presented in Section~\ref{subsec:weight-transform}. When the error vector $E(t)$ contains elements with opposite signs, we can ensure that the uncertain parameter vector $\mathbf{m}$ remains within its prescribed bounds in the parameter uncertainty space $M$ defined in Definition~\ref{def:vector}. This verification becomes crucial before applying the parameter-tying methodology or proceeding with the transformation to canonical form, as it validates that the plant has not violated its predefined uncertainty boundaries.

\section{Application to Vehicle Dynamics: State Estimation in Coupled Lateral-Roll Motion}
Vehicle state estimation typically relies on model-based observers such as Kalman Filter variants \cite{antonov2011unscented,ma2017state}, which provide optimal state predictions under certain conditions. These traditional approaches perform well when system models are accurately defined; however, they face significant limitations when dealing with uncertain parameters or unmeasurable inputs. The integrated lateral and roll dynamics present a particularly challenging scenario, as they require precise knowledge of road geometry effects including road bank angle ($\phi_r$) and road grade angle ($\theta_r$)—parameters that are difficult or prohibitively expensive to measure directly in production vehicles \cite{hajiloo2022integrated}. The Multi-model adaptive system theory developed in this paper provides an elegant solution to this challenge by offering a computationally efficient framework that can effectively handle parameter uncertainties without requiring direct measurement of road geometry. This approach enables robust state estimation even under varying road conditions while maintaining a minimal set of identification models.

\subsection{Vehicle Dynamic Model}
The vehicle dynamics model is developed under several key assumptions: the suspension system is represented by a torsional spring and damper; roll center height remains constant; small slip angle conditions apply to ensure linear tire behavior; longitudinal forces are small enough not to significantly affect lateral dynamics; and the model is linearized around straight-line driving. The operating point conditions assume zero lateral forces ($\bar{F}_{yf} = \bar{F}_{yr} = 0$), zero slip angles ($\bar{\alpha}_f = \bar{\alpha}_r = 0$), and zero steering angle ($\bar{\delta} = 0$). Small angle approximations are used for steering angle ($\sin \delta \approx \delta$, $\cos \delta \approx 1$), and a small road bank angle ($\phi_r$) is assumed. The vehicle lateral velocity is related to sideslip angle by $v = u\beta$, where longitudinal velocity ($u$) is assumed constant over the prediction horizon.

The dynamic equations describing the vehicle's sideslip ($\beta$), yaw rate ($r$), and roll angle ($\phi$) are:

\begin{align}
    \dot{\beta} = {} & \frac{m_s h_s}{mu} \left( m_s g h_s \cos \phi_r \cos \theta_r - \frac{k_\phi \phi}{I_c} - \frac{c_\phi \dot{\phi}}{I_c} \right) - r \nonumber \\
    & + \frac{1}{mu} \left( 1 + \frac{m_s^2 h_s^2}{m I_c} \right) F_Y - \frac{g}{u} \sin \phi_r \cos \theta_r \label{eq80}
\end{align}

\begin{align}
    \dot{r} &= \frac{1}{I_z} \left( l_f F_{yf} \cos \delta - l_r F_{yr} + l_f F_{xf} \sin \delta + M_{DB} \right), \label{eq81}
\end{align}

\begin{align}
    \ddot{\phi} &= \frac{m_s g h_s \cos \phi_r \cos \theta_r - k_\phi \phi - c_\phi \dot{\phi}}{I_c} + \frac{m_s h_s}{m I_c} F_Y, \label{eq82}
\end{align}

where $F_Y = F_{yf} \cos \delta + F_{xf} \sin \delta + F_{yr}$ represents the total lateral force and $I_c = I_{x,s} - \frac{m_s^2 h_s^2}{m}$ is the composite roll moment of inertia. The parameters $m_s$ and $m$ denote the sprung and total mass; $I_z$ and $I_{x,s}$ are the yaw and roll inertia; $l_f$ and $l_r$ are the distances from CG to front and rear axles; $k_\phi$ and $c_\phi$ are the roll stiffness and damping; $\phi_r$ and $\theta_r$ are road bank and grade angles; $h_s$ is the CG-to-roll-center height; $M_{DB}$ is the differential braking moment; and $g$ is gravitational acceleration.

The lateral tire forces are linearized around their operating points according to:

\begin{equation}
    F_{yi} = \bar{F}_{yi} + \bar{C}_{\alpha i} (\alpha_i - \bar{\alpha}_i) \label{eq:83}
\end{equation}

with slip angles for front and rear axles given by:
\begin{equation}
    \alpha_f = \beta + \frac{l_f r}{u} - \delta, \quad \alpha_r = \beta - \frac{l_r r}{u} \label{eq:84}
\end{equation}

\subsection{State-Space Representation}\label{subsec:state-space}
The equations can be reformulated in a state-space form:
\begin{equation}
\dot{x}(t) = A x(t) + B d(t) \label{eq:state-space}
\end{equation}
where $x = [\beta, r, \phi, \dot{\phi}]^T$, $d = \delta$ is the steering input, and
\begin{gather*}
A = \begin{bmatrix}
    A_{11} & A_{12} & A_{13} & A_{14} \\
    A_{21} & A_{22} & 0 & 0 \\
    0 & 0 & 1 & 0 \\
    A_{41} & A_{42} & A_{43} & A_{44}
\end{bmatrix}, \quad
B = \begin{bmatrix}
    B_{1} \\ B_{2} \\ 0 \\ B_{4} 
\end{bmatrix},
\end{gather*}
\begin{align*}
A_{11} &= \left(1 + \frac{m_s^2 h_s^2}{m I_c}\right) \frac{C_{\alpha f} + C_{\alpha r}}{mu}, \\
A_{12} &= \left(1 + \frac{m_s^2 h_s^2}{m I_c}\right) \frac{l_f C_{\alpha f} - l_r C_{\alpha r}}{mu^2} - 1, \\
A_{13} &= \frac{m_s h_s (m_s g h_s \cos \phi_r \cos \theta_r - k_\phi)}{mu I_c}, \\
A_{14} &= -\frac{m_s h_s c_\phi}{mu I_c}, \quad A_{21} = \frac{l_f C_{\alpha f} - l_r C_{\alpha r}}{I_z}, \\
A_{22} &= \frac{l_f^2 C_{\alpha f} + l_r^2 C_{\alpha r}}{I_z u}, \quad A_{41} = \frac{m_s h_s (C_{\alpha f} + C_{\alpha r})}{mu I_c}, \\
A_{42} &= \frac{m_s h_s (l_f C_{\alpha f} - l_r C_{\alpha r})}{mu I_c}, \\
A_{43} &= \frac{m_s g h_s \cos \phi_r \cos \theta_r - k_\phi}{I_c}, \quad A_{44} = -\frac{c_\phi}{I_c}, \\
B_{1} &= -\frac{C_{\alpha f}}{mu}, \quad B_{2} = -\frac{l_f C_{\alpha f}}{I_z}, \quad B_{4} = -\frac{m_s h_s C_{\alpha f}}{mu I_c}.
\end{align*}

\subsection{Multi-model Structure}
\begin{figure}[!t]
\centering
\includegraphics[width=\columnwidth]{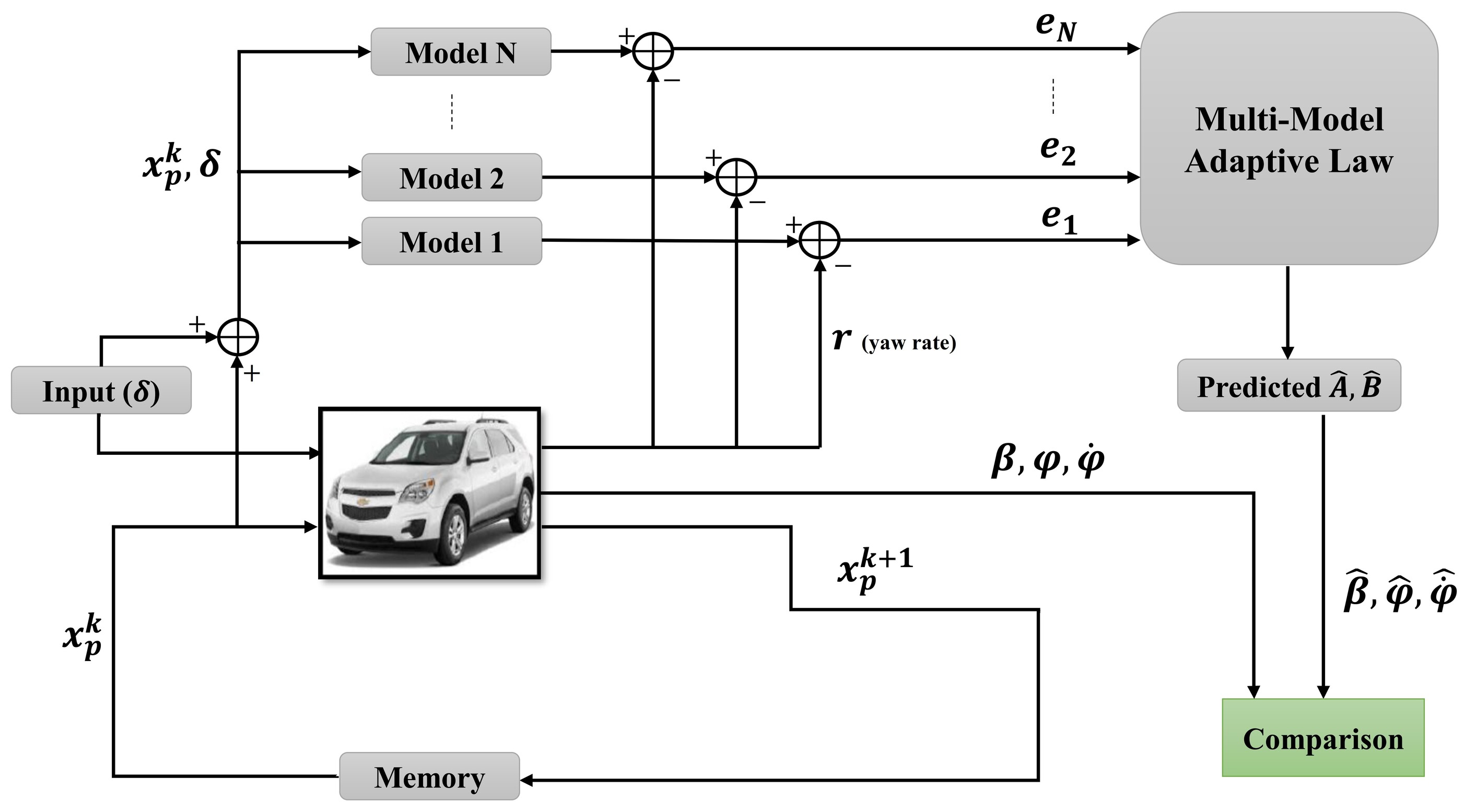}
\caption{Multi-model structure.}
\label{scheme}
\end{figure}

The overall structure of our proposed multi-model approach is illustrated in Fig. \ref{scheme}. The plant is a vehicle simulated in the CarSim environment, which receives steering input $\delta$ and produces yaw rate and system states as outputs. As detailed in Table~\ref{tab:param_variations}, the system accommodates parameter variations of ±30\% from their nominal values, along with specified variations in road conditions.

Traditional approaches to handling such uncertainty spaces would require $2^k$ vertices for $k$ uncertain parameters, resulting in 64 distinct models in our case when using constant bound values \cite{RN10}. However, we demonstrate that by leveraging the monotonicity properties established in Theorem~\ref{theorem:monotonicity}, this can be dramatically reduced to just two models.

To illustrate this reduction principle, consider the element $A_{13}$ introduced in Section~\ref{subsec:state-space}, specifically examining its behavior with respect to parameters $\theta_r$ and $k_\phi$. The element $A_{13}$ exhibits monotonic behavior with respect to both parameters within their defined bounds. For example, $A_{13}$ reaches its maximum value when $\cos(\theta_r)$ approaches one (corresponding to $\theta_r$ at its lower bound) and $k_\phi$ is at its lower bound. Consequently, a single model incorporating these lower bounds suffices to capture the maximum value of $A_{13}$.

This monotonicity principle extends to other parameters and system elements, allowing us to represent the entire uncertain space in controllable canonical form using only two models, as guaranteed by Theorem~\ref{theorem:monotonicity}. The effectiveness of this reduced representation will be validated through CarSim simulation results in the subsequent section.

\begin{table}[!t]
\caption{Parameter Variations from Nominal Values\label{tab:param_variations}}
\centering
\begin{tabular}{lccc}
\hline
Parameter & Lower Bound & Nominal Value & Upper Bound \\
\hline
$C_{af}$ (N/rad) & 56,280 & 80,400 & 104,520 \\
$C_{ar}$ (N/rad) & 57,890 & 82,700 & 107,510 \\
$k_{\phi}$ (N$\cdot$m/rad) & 25,200 & 36,000 & 46,800 \\
$c_{\phi}$ (N$\cdot$m$\cdot$s/rad) & 2,100 & 3,000 & 3,900 \\
$\phi_r$ (deg) & 0 & - & 4 \\
$\theta_r$ (deg) & 0 & - & 8 \\
\hline
\end{tabular}
\end{table}

\subsection{Simulation Results}
\begin{figure}[!t]
    \centering
    \begin{subfigure}{0.48\columnwidth}
        \centering
        \includegraphics[width=\textwidth]{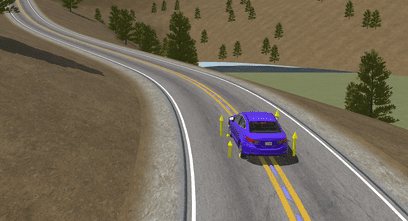}
        \caption{}
        \label{fig:carsim_V50}
    \end{subfigure}
    \hfill
    \begin{subfigure}{0.48\columnwidth}
        \centering
        \includegraphics[width=\textwidth]{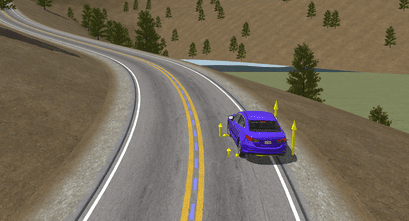}
        \caption{}
        \label{fig:carsim_V100}
    \end{subfigure}
    \caption{Simulated vehicle behavior in CarSim at constant speeds of: (a) 50 km/h and (b) 100 km/h.}
    \label{fig:carsim_combined}
\end{figure}

\begin{figure}[!t]
    \centering
    \begin{subfigure}{\columnwidth}
        \centering
        \includegraphics[width=\columnwidth]{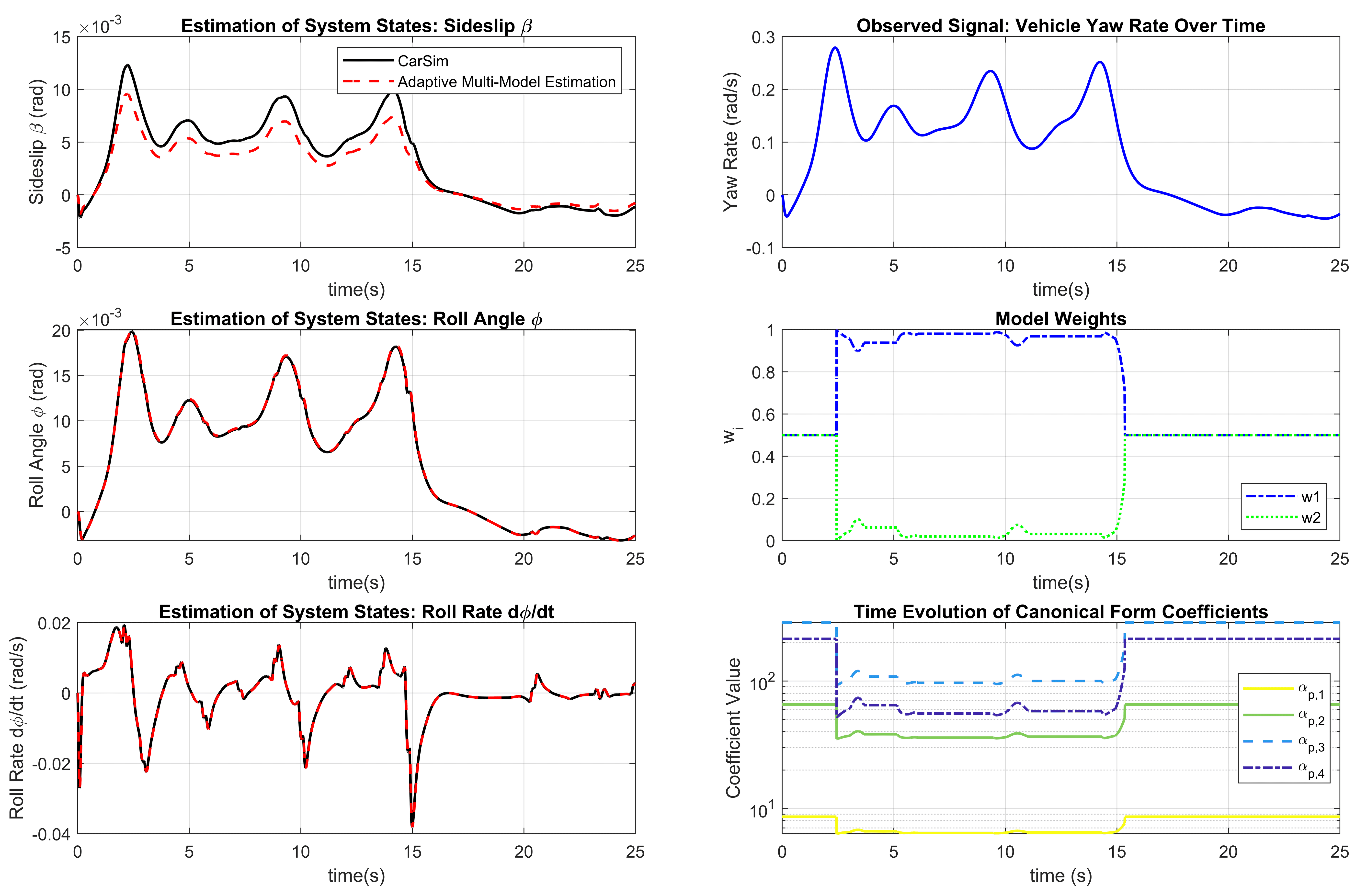}
        \caption{Simulation Results at Scenario 1: Target Speed of 50km/h}
        \label{subplots_V50}
    \end{subfigure}
    
    \vspace{1em}
    
    \begin{subfigure}{\columnwidth}
        \centering
        \includegraphics[width=\columnwidth]{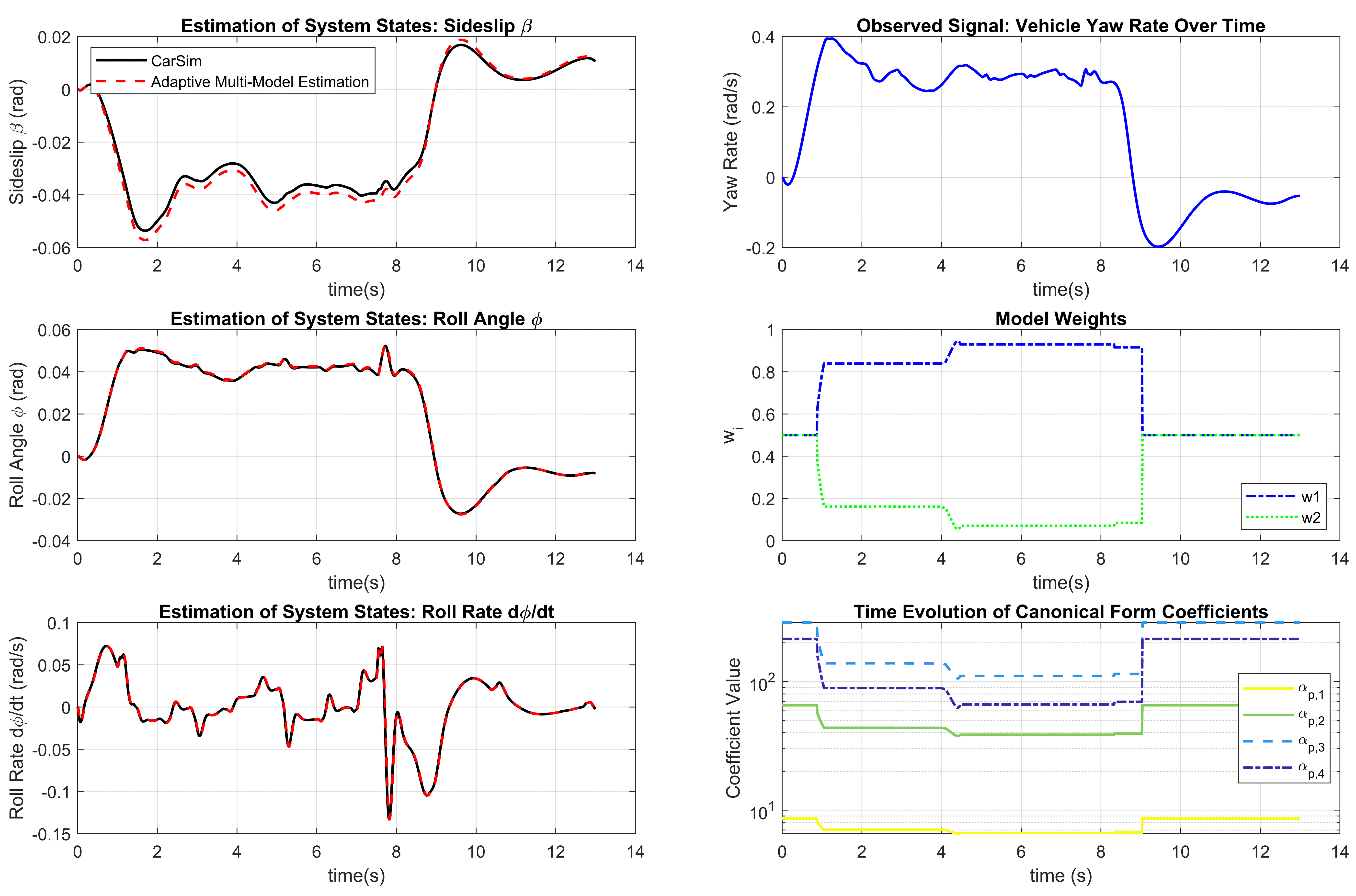}
        \caption{Simulation Results at Scenario 2: Target Speed of 100km/h}
        \label{subplots_V100}
    \end{subfigure}
    \caption{Simulation Results for Different Target Speeds}
    \label{fig:combined_results}
\end{figure}

The vehicle utilized in the simulations is a D-Class sedan, selected from the pre-built models offered in CarSim Ver 2022.0. Table \ref{tbl5} provides an outline of the principal properties of the simulated vehicle. Fig. \ref{fig:carsim_combined} shows the simulated vehicle behavior on a mountain road under different vehicle speed scenarios. Figs. \ref{subplots_V50} and \ref{subplots_V100} illustrate the simulation results for scenarios with target speeds of 50 km/h and 100 km/h, respectively.

In these scenarios, while the vehicle yaw rate serves as the primary observed signal, the multi-model adaptive law estimates three critical dynamic states: sideslip angle, roll angle, and roll rate. The estimation results demonstrate excellent accuracy for both roll rate and roll angle measurements. The sideslip angle estimation, while generally satisfactory, shows some deviation from the actual values. This discrepancy can be attributed primarily to the system linearization assumptions rather than limitations in the multi-model adaptive law itself, as the other estimated states maintain high accuracy under the same conditions.

\begin{table}[!t]
\caption{Vehicle Parameters Value\label{tbl5}}
\centering
\begin{tabular}{llc}
\hline
Symbol & Value & Description \\
\hline
$m$ & 1530 kg & Vehicle mass \\
$I_z$ & 2315.3 kgm\textsuperscript{2} & Mass moment of inertia \\
$I_\omega$ & 0.8 kgm\textsuperscript{2} & Wheel moment of inertia \\
$L$ & 2.78 m & Wheelbase \\
$t_f, t_r$ & 1.55 m & Front and rear track width \\
$R_{eff}$ & 0.325 m & Wheel effective radius \\
$a$ & 1.11 m & CG distance to front axle \\
$b$ & 1.67 m & CG distance to rear axle \\
$C_f$ & 80400 N/rad & Front tire cornering stiffness \\
$C_r$ & 82700 N/rad & Rear tire cornering stiffness \\
\hline
\end{tabular}
\end{table}

\section{Conclusion}
This paper has presented a novel theoretical framework for reducing the computational complexity of multi-model adaptive systems systems through systematic transformation to controllable canonical form. The key contributions of this work span both theoretical foundations and practical applications.

The Parameter-Tying Theorem introduced in this work provides a rigorous mathematical foundation for reducing the dimension of the uncertainty space while preserving system stability and performance. The theorem establishes that under specified monotonicity conditions, the extremal values of all matrix elements occur at the parameter space vertices, enabling a minimal selection of boundary values for constructing the parameter uncertainty vector. This framework is strengthened by the weight transformation analysis, which demonstrates that through appropriate similarity transformations, any system matrix within the parameter uncertainty space can be expressed as a convex combination of the vertex system matrices in companion form. This transformation preserves convexity properties while ensuring the resulting system maintains the companion form structure, complemented by explicit criteria for verifying plant inclusion within the convex hull of identification models through analysis of identification error signs across the model set.

The practical efficacy of this theoretical framework has been demonstrated through application to coupled lateral-roll vehicle dynamics. While traditional approaches would require $2^k$ vertices for $k$ uncertain parameters, resulting in 64 distinct models for our vehicle application, the Parameter-Tying Theorem enabled a dramatic reduction to just two models. The simulation results using CarSim validated this approach, showing excellent agreement between estimated and actual states across different operating conditions, particularly for roll rate and roll angle measurements.

This significant reduction in computational complexity while maintaining estimation accuracy demonstrates the practical value of our theoretical contributions. The framework's ability to handle high-dimensional systems with coupled dynamics, as demonstrated in the vehicle application, suggests broad applicability in scenarios where real-time implementation is crucial.

\appendices
\section{Detailed Analysis of Characteristic Polynomial Coefficients Bounds}
\label{app:char-poly}

This appendix provides a comprehensive analysis of the coefficients in the characteristic polynomial of matrix $A$. The analysis establishes bounds for each coefficient based on the element-wise bounds of the matrix.

\subsection{Analysis of $(n-1)$-degree Coefficient}
\label{app:n-1-coef}
The coefficient $c_{n-1}$ is equal to the negative trace of matrix $A$ \cite{horn2012matrix}:
\begin{equation}
c_{n-1} = -\text{tr}(A) = -\sum_{i=1}^n a_{ii}
\label{eq:app2}
\end{equation}

\begin{proposition}
Given the element-wise bounds on $A$, the coefficient $c_{n-1}$ is bounded as:
\begin{equation}
-\sum_{i=1}^n ub_{ii} \leq c_{n-1} \leq -\sum_{i=1}^n lb_{ii}
\label{eq:app3}
\end{equation}
\end{proposition}

\begin{proof}
Since $lb_{ii} \leq a_{ii} \leq ub_{ii}$ for each $i$:
\begin{enumerate}
    \item $-ub_{ii} \leq -a_{ii} \leq -lb_{ii}$ for each $i$
    \item Summing over all $i$ preserves the inequalities
    \item Therefore, $-\sum_{i=1}^n ub_{ii} \leq -\sum_{i=1}^n a_{ii} \leq -\sum_{i=1}^n lb_{ii}$
\end{enumerate}
\end{proof}

\subsection{Analysis of Constant Term $c_0$}
\label{app:constant-term}

The constant term $c_0$ is related to the determinant of $A$ \cite{horn2012matrix}:
\begin{equation}
c_0 = (-1)^n \det(A)
\label{eq:app4}
\end{equation}

\begin{definition}[Leibniz Determinant Formula]
For an $n \times n$ matrix $A$ and a permutation $\sigma \in S_n$, define:
\begin{equation}
P_\sigma(A) = \prod_{i=1}^n a_{i,\sigma(i)}
\label{eq:app9}
\end{equation}
where $S_n$ is the symmetric group of degree $n$. The determinant of $A$ is given by:
\begin{equation}
\det(A) = \sum_{\sigma \in S_n} \text{sgn}(\sigma) P_\sigma(A)
\label{eq:app10}
\end{equation}
where $\text{sgn}(\sigma)$ is the signature of permutation $\sigma$.
\end{definition}

\begin{proposition}
For an $n \times n$ matrix $A$ with entries bounded by $lb_{ij} \leq a_{ij} \leq ub_{ij}$, the constant term $c_0$ of its characteristic polynomial is bounded as follows:

For $n$ even:
\begin{equation}
\begin{aligned}
lb_{c_0} &= \sum_{\sigma: \text{sgn}(\sigma)=1} \prod_{i=1}^n lb_{i,\sigma(i)} + \sum_{\sigma: \text{sgn}(\sigma)=-1} (-1)\prod_{i=1}^n ub_{i,\sigma(i)} \\
ub_{c_0} &= \sum_{\sigma: \text{sgn}(\sigma)=1} \prod_{i=1}^n ub_{i,\sigma(i)} + \sum_{\sigma: \text{sgn}(\sigma)=-1} (-1)\prod_{i=1}^n lb_{i,\sigma(i)}
\end{aligned}
\label{eq:app11}
\end{equation}

For $n$ odd:
\begin{equation}
\begin{aligned}
lb_{c_0} &= -\left(\sum_{\sigma: \text{sgn}(\sigma)=1} \prod_{i=1}^n ub_{i,\sigma(i)} + \sum_{\sigma: \text{sgn}(\sigma)=-1} (-1)\prod_{i=1}^n lb_{i,\sigma(i)}\right) \\
ub_{c_0} &= -\left(\sum_{\sigma: \text{sgn}(\sigma)=1} \prod_{i=1}^n lb_{i,\sigma(i)} + \sum_{\sigma: \text{sgn}(\sigma)=-1} (-1)\prod_{i=1}^n ub_{i,\sigma(i)}\right)
\end{aligned}
\label{eq:app12}
\end{equation}
\end{proposition}

\begin{proof}
The proof follows these key steps:

\begin{enumerate}
    \item From $c_0 = (-1)^n \det(A)$ and the Leibniz formula:
    \begin{equation}
    \det(A) = \sum_{\sigma \in S_n} \text{sgn}(\sigma) \prod_{i=1}^n a_{i,\sigma(i)}
    \label{eq:app13}
    \end{equation}

    \item For each $a_{ij}$:
    \begin{equation}
    lb_{ij} \leq a_{ij} \leq ub_{ij}
    \label{eq:app14}
    \end{equation}

    \item For any $\sigma \in S_n$:
    \begin{itemize}
        \item When $\text{sgn}(\sigma) = 1$:
        \begin{equation}
        \prod_{i=1}^n lb_{i,\sigma(i)} \leq \text{sgn}(\sigma)\prod_{i=1}^n a_{i,\sigma(i)} \leq \prod_{i=1}^n ub_{i,\sigma(i)}
        \label{eq:app15}
        \end{equation}
        
        \item When $\text{sgn}(\sigma) = -1$:
        \begin{equation}
        -\prod_{i=1}^n ub_{i,\sigma(i)} \leq \text{sgn}(\sigma)\prod_{i=1}^n a_{i,\sigma(i)} \leq -\prod_{i=1}^n lb_{i,\sigma(i)}
        \label{eq:app16}
        \end{equation}
    \end{itemize}

    \item The factor $(-1)^n$ affects the direction of inequalities:
    \begin{itemize}
        \item When $n$ is even: inequalities preserve direction
        \item When $n$ is odd: inequalities reverse direction
    \end{itemize}

    \item This yields the stated bounds in equations \eqref{eq:app11} and \eqref{eq:app12}.
\end{enumerate}
\end{proof}

\subsection{Analysis of Intermediate Coefficients}
\label{app:intermediate-coef}

\begin{definition}[Principal Submatrix]
For $S \subseteq \{1,\ldots,n\}$, the principal submatrix $A_S$ is formed by rows and columns of $A$ indexed by $S$.
\end{definition}

For $1 \leq k \leq n-2$, coefficient $c_k$ is given by \cite{horn2012matrix}:
\begin{equation}
c_k = (-1)^{n-k} \sum_{S \in \mathcal{C}(n,n-k)} \det(A_S)
\label{eq:app17}
\end{equation}
where $\mathcal{C}(n,n-k)$ is the set of all $(n-k)$-combinations of $\{1,\ldots,n\}$.

The intermediate coefficients $c_k$ are related to sums of principal minors. For each coefficient $c_k$, we need to compute the determinant of all submatrices of size $(n-k)$. The lower bound for each minor is obtained by setting each element of the submatrix to its lower bound, and the upper bound by setting each element to its upper bound.

Thus, for each coefficient $c_k$, the bounds can be expressed as:
\begin{equation}
lb_{c_k} \leq c_k \leq ub_{c_k}
\label{eq:app18}
\end{equation}
where:
\begin{equation}
\begin{aligned}
lb_{c_k} &= (-1)^{n-k} \sum_{S \in C(n,n-k)} \min(\det(A_S)) \\
ub_{c_k} &= (-1)^{n-k} \sum_{S \in C(n,n-k)} \max(\det(A_S))
\end{aligned}
\label{eq:app19}
\end{equation}

Here, $C(n,n-k)$ is the set of all subsets of $\{1, \ldots, n\}$ of size $n-k$, and $A_S$ is the principal submatrix of $A$ formed by rows and columns in subset $S$. The determinant bounding procedure outlined in the previous subsection can similarly be applied to each principal submatrix $A_S$, establishing that $c_k$ is constrained by the lower and upper bounds of the elements $a_{ij}$.

\section{Weight Transformation Analysis}\label{app:weight-transform}

This appendix provides a comprehensive derivation of the weight transformation relationships for multi-model systems in canonical form. The analysis establishes how convex combinations are preserved under coordinate transformations and proves the properties of the transformed weights.

\subsection{Preliminaries}

For each vertex system $i$, there exists a transformation matrix $T_i \in \mathbb{R}^{n \times n}$ that transforms the system to its controllable canonical form:
\begin{equation}
\label{eq35}
\begin{aligned}
A_i &= T_i^{-1} \bar{A}_{i} T_i \\
B_i &= T_i^{-1} \bar{B}_{i}
\end{aligned}
\end{equation}
where $\bar{A}_{i} \in \mathbb{R}^{n \times n}$ and $\bar{B}_{i} \in \mathbb{R}^{n \times h}$ are the system matrices in canonical form.

Similarly, for the uncertain system:
\begin{equation}
\label{eq36}
\begin{aligned}
A_p &= T_p^{-1} \bar{A}_{p} T_p \\
B_p &= T_p^{-1} \bar{B}_{p}
\end{aligned}
\end{equation}

The multi-model assumption states that:
\begin{equation}
\label{eq33}
A_p = \sum_{i=1}^{N} w_iA_i,\quad B_p = \sum_{i=1}^N w_iB_i
\end{equation}
with convexity conditions:
\begin{equation}
\label{eq34}
\sum_{i=1}^{N} w_i = 1,\quad w_i \geq 0,\quad \forall i \in \{1,\dots,N\}
\end{equation}

\subsection{Derivation of Weight Transformation}

\noindent\textbf{Step 1: Substituting Multi-Model Representation}

We begin by substituting the multi-model assumption for $[A_p, B_p]$ into the equation \ref{eq5} for the uncertain system in canonical form:
\begin{equation}
\label{eq37}
\begin{aligned}
T_p^{-1} \bar{A}_{p} T_p &= \sum_{i=1}^{N} w_i A_i \\
T_p^{-1} \bar{B}_{p} &= \sum_{i=1}^{N} w_i B_i
\end{aligned}
\end{equation}

Next, substitute the controllable canonical form representation for each vertex system $i$:
\begin{equation}
\label{eq38}
\begin{aligned}
T_p^{-1} \bar{A}_{p} T_p &= \sum_{i=1}^{N} w_i T_i^{-1} \bar{A}_{i} T_i \\
T_p^{-1} \bar{B}_{p} &= \sum_{i=1}^{N} w_iT_i^{-1} \bar{B}_{i}
\end{aligned}
\end{equation}

\noindent\textbf{Step 2: Rearranging Terms}

Rearrange to isolate $\bar{A}_{p}$ and $\bar{B}_{p}$ on the left-hand side:
\begin{equation}
\label{eq39}
\begin{aligned}
\bar{A}_{p} &= T_p \left( \sum_{i=1}^{N}w_i T_i^{-1} \bar{A}_{i} T_i \right) T_p^{-1} \\
\bar{B}_{p} &= T_p \left( \sum_{i=1}^{N}w_i T_i^{-1} \bar{B}_{i}\right)
\end{aligned}
\end{equation}

Using the distributive property of matrix multiplication:
\begin{equation}
\label{eq40}
\begin{aligned}
\bar{A}_{p} &= \sum_{i=1}^{N} w_i \left( T_p T_i^{-1} \bar{A}_{i} T_i T_p^{-1} \right) \\
\bar{B}_{p} &= \sum_{i=1}^{N} w_i \left( T_p T_i^{-1} \bar{B}_{i} \right)
\end{aligned}
\end{equation}

\noindent\textbf{Step 3: Introduction of Similarity Transformation}

Define the similarity transformation matrix:
\begin{equation}
S_i = T_p T_i^{-1} \in \mathbb{R}^{n \times n}
\end{equation}

This allows us to write:
\begin{equation}
\label{eq41}
\begin{aligned}
\bar{A}_{p} &= \sum_{i=1}^{N} w_i \left( S_i \bar{A}_{i} S_i^{-1} \right) \\
\bar{B}_{p} &= \sum_{i=1}^{N} w_iS_i \bar{B}_{i}
\end{aligned}
\end{equation}

\noindent\textbf{Step 4: Definition of Matrix-Valued Weights}

Define the matrix-valued weights:
\begin{equation}
\tilde{w}_i = w_i S_i \in \mathbb{R}^{n \times n}
\end{equation}

Then:
\begin{equation}
\label{eq42}
\begin{aligned}
\bar{A}_p &= \sum_{i=1}^{N} w_i \left( S_i \bar{A}_i S_i^{-1} \right) = \sum_{i=1}^{N} (w_i S_i) \bar{A}_i S_i^{-1} \\
&= \sum_{i=1}^{N} \tilde{w}_i \bar{A}_i S_i^{-1}
\end{aligned}
\end{equation}

Similarly for $\bar{B}_p$:
\begin{equation}
\label{eq43}
\begin{aligned}
\bar{B}_p &= \sum_{i=1}^{N} w_i S_i \bar{B}_i = \sum_{i=1}^{N} \tilde{w}_i \bar{B}_i
\end{aligned}
\end{equation}

\subsection{Properties of Transformed Weights}

\noindent\textbf{Sum-to-One Property}

Using the fact that in controllable canonical form:
\begin{equation}
\bar{B}_p=\tilde{B}_i=\begin{bmatrix} 0 & 0 & \cdots & 1 \end{bmatrix}^T
\end{equation}

We have:
\begin{equation}
\begin{bmatrix} 0 & 0 & \cdots & 1 \end{bmatrix}^T = \sum_{i=1}^{N} \tilde{w}_i\tilde{B}_i = \sum_{i=1}^{N} \tilde{w}_i \begin{bmatrix} 0 & 0 & \cdots & 1 \end{bmatrix}^T
\end{equation}

Therefore:
\begin{equation}
\sum_{i=1}^{N} \tilde{w}_i = 1
\end{equation}

\noindent\textbf{Non-Negativity}

Since $w_i \geq 0$ for all $i$, and $S_i$ are similarity transformation matrices (and therefore invertible), the matrix-valued weights $\tilde{w}_i = w_i S_i$ are non-zero matrices when $w_i > 0$.

\subsection{Explicit Form of Transformation Matrices}

By expanding the sum of new weights:
\begin{equation}
1 = \sum_{i=1}^{N} \tilde{w}_i = \sum_{i=1}^{N} w_i S_i = \sum_{i=1}^{N} w_i T_p T_i^{-1}
\end{equation}

Factoring out $T_p$:
\begin{equation}
1 = T_p \left( \sum_{i=1}^{N} w_i T_i^{-1} \right)
\end{equation}

Therefore:
\begin{equation}
T_p^{-1} = \sum_{i=1}^{N} w_i T_i^{-1}
\end{equation}

Taking the inverse:
\begin{equation}
T_p = \left( \sum_{i=1}^{N} w_i T_i^{-1} \right)^{-1}
\end{equation}

This gives the explicit form for $S_i$:
\begin{equation}
\label{eq51}
S_i = \left( \sum_{j=1}^{N} w_j T_j^{-1} \right)^{-1} T_i^{-1}
\end{equation}

\subsection{Final Form and Implementation}

Substituting the expression for $S_i$ back into the equation \ref{eq41} for $\bar{A}_p$:
\begin{equation}
\label{eq52}
\bar{A}_p = \sum_{i=1}^{N} w_i \left[ \left( \sum_{j=1}^{N} w_j T_j^{-1} \right)^{-1} T_i^{-1} \bar{A}_i T_i \left( \sum_{j=1}^{N} w_j T_j^{-1} \right) \right]
\end{equation}

This formulation establishes that the canonical form of the uncertain system is itself a convex combination of similarity-transformed vertex systems, with the transformation matrices dependent on the convex weights.

The practical implementation involves:
1. Determining scalar weights $w_i$ from plant identification errors in the original coordinate system
2. Computing transformation matrices using equation \eqref{eq51}
3. Applying the transformations for closed-loop analysis
4. Verifying that the transformed system maintains canonical form structure

This complete framework ensures preservation of both the multi-model structure and canonical form properties while enabling systematic analysis of uncertain systems.

\section{Proof of Plant Inclusion Criterion}\label{app:plant_inclusion}

This appendix provides the complete proof of Theorem~\ref{theorem:plant_inclusion}, which establishes the necessary and sufficient conditions for plant inclusion within the convex hull of identification models. The proof is divided into forward and backward directions, with each direction established through a sequence of rigorous mathematical arguments.

\noindent\textbf{Proof.} (forward direction) 

Given a vector \( E(t) = [e_1(t) \ e_2(t) \ \dots \ e_N(t)]\) with components of different signs, there exist a vector $w$ such that:
\begin{equation}
\sum_{i=1}^N w_i(t) e_i(t) = 0 
\label{eq59}
\end{equation} 
where \( w_i > 0 \) and \( \sum_{i=1}^N w_i = 1 \). To prove that, let's define the set \( A = \{z \in \mathbb{R}^n \mid z_i \geq 0, \sum_{i=1}^{n} z_i = 1\} \), which is a standard $n-1$ simplex. Now let's define the function $f: A \to \mathbb{R}$ as:
\begin{equation}
f(z) = E(t) \cdot z = \sum_{i=1}^n e_i(t) z_i
\label{eq60}
\end{equation}

Consider two points in $A$ as $p = (1,0,0,\ldots,0)$ and $q = (0,0,\ldots,0,1)$. Since $E(t)$ has components of different signs, without loss of generality, assume:
\begin{equation}
\begin{aligned}
f(p) &= e_1(t) > 0 \\
f(q) &= e_n(t) < 0
\end{aligned}
\label{eq61}
\end{equation}

According to the Bolzano's theorem, if a continuous function has values of opposite sign inside an interval, then it has a root in that interval. As function $f$ is continuous, it can be concluded that there exists a vector $w$ to satisfy the conditions in (\ref{eq59}).

Given the identification error \(e_i(t) = x_i(t) - x_p(t)\), the condition $\sum_{i=1}^N w_i(t) e_i(t) = 0$ can be rewritten as:
\begin{equation}
x_p(t) = \sum_{i=1}^N w_i(t) x_i(t)
\label{eq62}
\end{equation}
where \( w_i > 0 \) and \( \sum_{i=1}^N w_i = 1 \). By making the derivative from both sides:
\begin{equation}
\dot{x}_p(t) = \sum_{i=1}^N \left(\dot{w}_i(t) x_i(t) + w_i(t) \dot{x}_i(t)\right)
\label{eq63}
\end{equation}
By substituting equations \ref{eq1} and state equations for identification models into equation \ref{eq63}:
\begin{equation}
\begin{aligned}
A_p(\mathbf{m}) x_p(t) + B_p(\mathbf{m}) d(t) = &\sum_{i=1}^N \left(w_i(t)A_i + \dot{w}_i(t)\right)x_i(t)\\
&+ d(t)\sum_{i=1}^N w_i(t)B_i
\end{aligned}
\label{eq64}
\end{equation}
By arranging the expressions containing the input d(t) and other remaining terms, we have:
\begin{equation}
B_p(\mathbf{m}) = \sum_{i=1}^N w_i(t)B_i
\label{eq65}
\end{equation}
\vspace{-10pt}
\begin{equation}
A_p(\mathbf{m}) x_p(t) = \sum_{i=1}^N \left(w_i(t)A_i + \dot{w}_i(t)\right)x_i(t)
\label{eq66}
\end{equation}
By substituting equation \ref{eq62} into equation \ref{eq66}:
\begin{equation}
\sum_{i=1}^N \left(w_i(t)A_i - w_i(t)A_p(\mathbf{m}) + \dot{w}_i(t)\right)x_i(t)=0
\label{eq67}
\end{equation}

The state vectors $x_i(t)$ are linearly independent. To prove that, consider the linear combination of the state vectors:  
\begin{equation}
\sum_{i=1}^N c_i x_i(t) = 0
\label{eq68}
\end{equation}
where \( c_i(s) \) are constants. We want to prove that \( c_i=0 \) for all $i$. By contradiction, assume that there exists a non-trivial solution where at least one $c_i \neq 0$. Without loss of generality, let $c_1 \neq 0$, which would imply that $x_1(t)$ is a linear combination of the other state vectors, but each identification model provides a unique solution $x_i(t)$, leading to a contradiction. Therefore, the only solution to the linear combination of equation \ref{eq68} must be \( c_i=0 \) for all $i$.

Now since the vectors $x_i(t)$ are linearly independent, according to equation \ref{eq67} for all $i$: 
\begin{equation}
w_i(t)A_p(\mathbf{m}) = w_i(t)A_i + \dot{w}_i(t)
\label{eq69}
\end{equation}
By summation of both sides of equations in the form \ref{eq69} for $i = 1, \dots, N$: 

\begin{equation}
\begin{aligned}
&w_1(t) A_p(\mathbf{m}) + \cdots + w_N(t) A_p(\mathbf{m}) = \\
&\qquad \left(w_1(t) A_1 + \cdots + w_N(t) A_N \right) + \left( \dot{w}_1(t) + \cdots + \dot{w}_N(t) \right)
\end{aligned}
\label{eq70}
\end{equation}
In compact form:
\begin{equation}
\sum_{i=1}^{N} w_i(t) A_p(\mathbf{m}) = \sum_{i=1}^{N} w_i(t) A_i + \sum_{i=1}^{N} \dot{w}_i(t)
\label{eq71}
\end{equation}

Since \( \sum_{i=1}^N w_i = 1 \), the summation of derivatives of $w_i(t)$ becomes zero. Therefore, the equation \ref{eq71} will be simplified as:
\begin{equation}
A_p(\mathbf{m}) = \sum_{i=1}^{N} w_i(t) A_i
\label{eq72}
\end{equation}

The combination of equations \ref{eq65} and \ref{eq72} subject to constraints \( w_i > 0 \) and \( \sum_{i=1}^N w_i = 1 \) shows that the plant matrices $A_p(\mathbf{m})$ and $B_p(\mathbf{m})$ lie within the convex hull formed by the set $[A_i \ B_i]: i = 1, \dots, N$, which completes the proof.
\\

\noindent\textbf{Proof.} (backward direction)

Given the fact that the plant matrices $A_p(\mathbf{m})$ and $B_p(\mathbf{m})$ lie within the convex hull formed by the set $[A_i \ B_i]: i = 1, \dots, N$, the convexity property can be re-expressed as:
\begin{equation}
A_p(\mathbf{m}) x_p(t) + B_p(\mathbf{m}) d(t) = \sum_{i=1}^{N} w_i \left[ A_i x_p(t) + B_i d(t) \right]
\label{eq73}
\end{equation}
where \( w_i > 0 \) and \( \sum_{i=1}^N w_i = 1 \). By the fact that all identification models have initial conditions $x_i(t_0)=x_p(t_0)$, at $t=t_0$, the equation \ref{eq73} will become:
\begin{equation}
A_p(\mathbf{m}) x_p(t_0) + B_p(\mathbf{m}) d(t_0) = \sum_{i=1}^{N} w_i \left[ A_i x_i(t_0) + B_i d(t_0) \right]
\label{eq74}
\end{equation}
By substituting (\ref{eq1}) into (\ref{eq74}):
\begin{equation}
\dot{x}_p(t_0)=\sum_{i=1}^{N} w_i \dot{x}_i(t_0)
\label{eq75}
\end{equation}
Integrating (\ref{eq75}) results in:
\begin{equation}
x_p(t)-x_p(t_0)=\sum_{i=1}^{N} w_i\left[ x_i(t)-x_i(t_0) \right]
\label{eq76}
\end{equation}
Which will be simplified to:
\begin{equation}
x_p(t)=\sum_{i=1}^{N} w_ix_i(t)
\label{eq77}
\end{equation}
Given \( \sum_{i=1}^N w_i = 1 \), equation \ref{eq77} can be rearranged as: 
\begin{equation}
\sum_{i=1}^{N} w_i\left[x_p(t)-x_i(t) \right]=0
\label{eq78}
\end{equation}

Using the definition of identification error \(e_i(t) = x_i(t) - x_p(t)\), (\ref{eq78}) can be re-expressed in matrix form as:
\begin{equation}
\begin{aligned}
&E(t)W = 0, \quad E(t) = \begin{bmatrix} e_1(t) \ e_2(t) \ \dots \ e_N(t) \end{bmatrix},\\ &W^T = \begin{bmatrix} w_1 \ w_2 \ \dots \ w_N \end{bmatrix}
\end{aligned}
\label{eq79}
\end{equation}

At this stage, we will prove that there exist at least two elements of \( E(t) \) with opposite signs. By contradiction, suppose all elements of \( E(t) \) have the same sign, meaning either $e_i(t) \geq 0$, or $e_i(t) \leq 0$ for all $i$. Given that \( w_i > 0 \) for all $i$, then each product $e_i(t)w_i$ will be non-negative (or non-positive). In order to satisfy (\ref{eq79}), each single term $e_i(t)w_i$ should be zero, which results in $e_i(t)=0$ for all $i$. However, this contradicts the assumption that $E(t)$ has non-zero elements, completing the proof. ~$\square$

\vspace{\baselineskip}  % Adds one line of space

\setlength\intextsep{0pt} % align top of photo with text
\begin{wrapfigure}{l}{0.13\textwidth}
    \centering
    \includegraphics[width=0.15\textwidth]{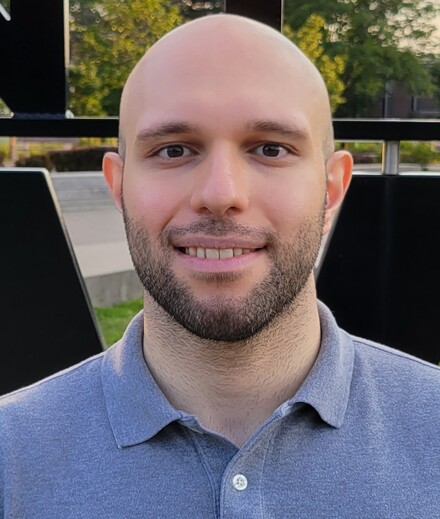}
\end{wrapfigure}
\noindent {\small \textbf{Farid Mafi} received the B.Sc. and M.Sc. degrees in mechanical engineering from the Sharif University of Technology, Tehran, Iran in 2017 and 2019, respectively. He is currently a Ph.D. Candidate at the University of Waterloo, specializing in vehicle control systems. His research interests include vehicle dynamics, control, estimation, and optimization.}\\

\setlength\intextsep{0pt} % align top of photo with text
\begin{wrapfigure}{l}{0.13\textwidth}
    \centering
    \includegraphics[width=0.15\textwidth]{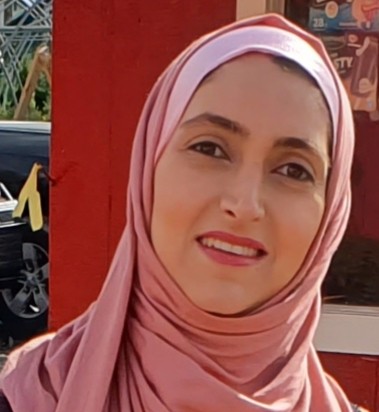}
\end{wrapfigure}
\noindent {\small \textbf{Ladan Khoshnevisan} (Member, IEEE) received her B.S. in electrical engineering from the University of Science and Technology, Tehran, in 2008, and her M.Sc. and Ph.D. from Tarbiat Modares University and the University of Tehran in 2010 and 2017, respectively. In 2017, she was an International Visiting Ph.D. Student at the University of Waterloo and later became a research assistant in Mechanical and Mechatronics Engineering. From 2022 to 2024, she was a Post-Doctoral Fellow in Applied Mathematics and is currently a Post-Doctoral Fellow in Mechanical and Mechatronics Engineering at the University of Waterloo. Her research interests include control systems, cyber security, and fault-tolerant control. }\\

\setlength\intextsep{0pt} % align top of photo with text
\begin{wrapfigure}{l}{0.13\textwidth}
    \centering
    \includegraphics[width=0.15\textwidth]{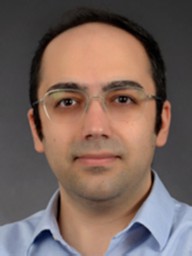}
\end{wrapfigure}
\noindent {\small \textbf{Mohammad Pirani} (Senior Member, IEEE) received the M.A.Sc. degree in electrical and computer engineering and a Ph.D. degree in mechanical and mechatronics engineering from the University of Waterloo, in 2014 and 2017, respectively. He is an Assistant Professor at the Department of Mechanical Engineering, University of Ottawa, Canada. He was a Research Assistant Professor with the Department of Mechanical and Mechatronics Engineering, University of Waterloo, from 2022 to 2023. He held post-doctoral researcher positions with the University of Toronto, from 2019 to 2021, and the KTH Royal Institute of Technology, Sweden, from 2018 to 2019. His research interests include resilient and fault-tolerant control, networked control systems, and multi-agent systems.}\\

\setlength\intextsep{0pt} % align top of photo with text
\begin{wrapfigure}{l}{0.13\textwidth}
    \centering
    \includegraphics[width=0.15\textwidth]{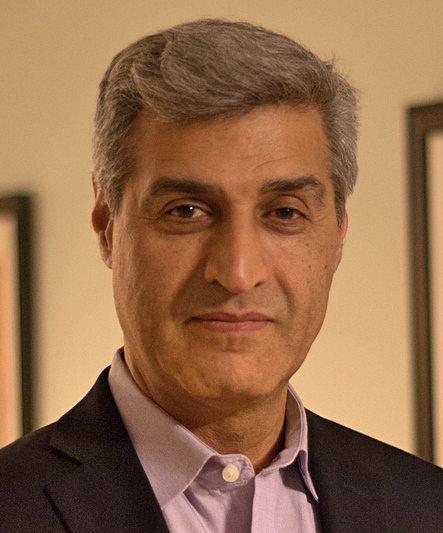}
\end{wrapfigure}
\noindent {\small \textbf{Amir Khajepour} (Senior Member, IEEE)  is a professor of Mechanical and Mechatronics Engineering at the University of Waterloo. He held the Tier 1 Canada Research Chair in Mechatronic Vehicle Systems from 2008 to 2022 and the Senior NSERC/General Motors Industrial Research Chair in Holistic Vehicle Control from 2017 to 2022. His work has resulted in the training of over 160 PhD and MASc students, 33 patents, over 600 research papers, numerous technology transfers, and several start-up companies. He has been recognized with the Engineering Medal from Professional Engineering Ontario and is a fellow of the Engineering Institute of Canada, the American Society of Mechanical Engineering, and the Canadian Society of Mechanical Engineering.}

\end{document}